# Simplification of the tug-of-war model for cellular transport in cells

Yunxin Zhang[*]

*Shanghai Key Laboratory for Contemporary Applied Mathematics,*

*Centre for Computational System Biology,*

*School of Mathematical Sciences, Fudan University, Shanghai 200433, China.*

(Dated: June 19, 2010)

The transport of organelles and vesicles in living cells can be well described by a kinetic tug-of-war model advanced by Müller, Klumpp and Lipowsky. In which, the cargo is attached by two motor species, kinesin and dynein, and the direction of motion is determined by the number of motors which bind to the track. In recent work [Phys. Rev. E **79**, 061918 (2009)], this model was studied by mean field theory, and it was found that, usually the tug-of-war model has one, two, or three distinct stable stationary points. However, the results there are mostly obtained by numerical calculations, since it is hard to do detailed theoretical studies to a two-dimensional nonlinear system. In this paper, we will carry out further detailed analysis about this model, and try to find more properties theoretically. Firstly, the tug-of-war model is simplified to a one-dimensional equation. Then we claim that the stationary points of the tug-of-war model correspond to the roots of the simplified equation, and the stable stationary points correspond to the roots with positive derivative. Bifurcation occurs at the corresponding parameters, under which the simplified one-dimensional equation exists root with zero derivative. Using the simplified equation, not only more properties of the tug-of-war model can be obtained analytically, the related numerical calculations will become more accurate and more efficient. This simplification will be helpful to future studies of the tug-of-war model.

I. INTRODUCTION

In live cells, the organelles and vesicle are usually transported along microtubule by two species of motor proteins, kinesins and dyneins [1–6]. In which, kinesin tries to move the cargo to the plus end of the microtubule, and dynein tries to move to the minus end [7–13]. To understand the mechanism of such transport, a tug-of-war model has been devised by Lipowsky and coworkers in [3, 4]. Recently, verified by the Monte Carlo simulation, the mean field method is employed by the author to analyze this model [14]. The results established that the tug-of-war model may realize states with either one, two, or three distinct stable stationary modes of motion, which depend on parameters of the two types of motor protein and on the ratio $\nu = N_+/N_-$, of the available number $N_+$ of kinesin to the number $N_-$ of dynein [15]. However, most of the results obtained in both [14] and [15] are based on numerical calculations, since it is really hard to get valuable results only by theoretical analysis.

The main difficulty in the mathematical analysis of tug-of-model is that, this model is composed of two nonlinear ordinary different equations. So, at steady state, it is a two-dimensional nonlinear system. In this paper, by variable substitution, this two-dimensional system is simplified into a one-dimensional equation firstly. Then we proved that the stationary states of the tug-of-war model correspond to the roots of this one-dimensional equation, and their stabilities are determined by the sign of derivative at the corresponding roots. For more details, stable stationary states correspond to roots with positive derivative, unstable stationary states correspond to roots with negative derivative, and bifurcation occurs at the model parameters under which the simplified equation exists root with zero derivative. The advantage of this simplification is that, not only it makes the model theoretically tractable,

---
∗Electronic address: `xyz@fudan.edu.cn`

but also it is helpful to simplify the numerical calculations.

The organization of this paper is as follows. The tug-of-war model and its mean field limit will be briefly introduced in the next Section, then in Section **III**, we will simplify the two-dimensional system into a one-dimensional equation, and discuss the relations between them. Finally, concluding remarks will be given in the Section **IV**.

## II. TUG-OF-WAR MODEL AND MEAN FIELD LIMIT

In tug-of-war model, the cargo is assumed to be bound by $N_+$ kinesins and $N_-$ dyneins. Whereas $N_+$ and $N_-$ will be supposed fixed, the numbers $n_+(t)$ of kinesin and $n_-(t)$ of dynein that attach the cargo to the track vary stochastically. The corresponding rates are specified by the properties of individual motors. To be specific regarding the motor parameters, (i) the attachment or "on" rates $k_+^{\text{on}}$ and $k_-^{\text{on}}$ are postulated to be load independent, and (ii) detachment or "off" rates $k_+^{\text{off}}(F)$ and $k_-^{\text{off}}(F)$ that are supposed to depend on the load force, $F$, felt by the individual motor according to

$$k_\pm^{\text{off}}(F) = k_{0\pm}^{\text{off}} \exp(|F|/F_d^\pm), \tag{1}$$

where the detachment forces, $F_d^+$ and $F_d^-$, characterize the individual $+$ and $-$ motors. (iii) The load-speed relations for single $+$ and $-$ motor proteins, i.e., kinesin and dynein, in the tug-of-war model are taken to obey the piecewise linear relations

$$V_\pm(F) = V_F^\pm[1 - (F/F_S^\pm)] \quad \text{for} \quad 0 \leq F \leq F_S^\pm, \tag{2a}$$
$$= V_B^\pm[1 - (F/F_S^\pm)] \quad \text{for} \quad F \geq F_S^\pm, \tag{2b}$$
$$= V_F^\pm \quad \text{for} \quad F \leq 0, \tag{2c}$$

with corresponding zero-load speeds $V_F^\pm$, superstall speed amplitude $V_B^\pm$, and stall forces $F_S^\pm$. It is supposed (iv) that all the individual motor velocities, $V_+$ and $V_-$, match the cargo velocity $V_C$. Finally, for simplicity, the model assumes (v) that the instantaneous load is shared *equally* between the $n_+(t)$ plus-end directed motors that happen to be attached to the track at time $t$ and, likewise for the $n_-(t)$ attached plus-end directed motors.

The model assumptions (i)–(v) now imply

$$V_C(n_+, n_-) = V_+(F_C/n_+) = -V_-(-F_C/n_-), \tag{3}$$

where $F_C$ denotes the load on the cargo defined by

$$F_C := n_+ F_+ = -n_- F_- + F_{ext}, \tag{4}$$

where $F_+$ and $F_-$ are the loads felt by each individual $+$ and $-$ motor respectively, while $F_{ext}$ denotes an externally imposed load on the cargo; all the loads are defined to be positive when directed towards the minus end of the track. For simplicity, it is assumed in this paper that the external load $F_{\text{ext}}$ vanishes. The independence of the individual motors and their shared load implies the overall detachment rates

$$\epsilon_\pm(n_+, n_-) = n_\pm k_\pm^{\text{off}}(n_+, n_-) = k_{0\pm}^{\text{off}} n_\pm \exp[|F_C|/n_\pm F_d^\pm], \tag{5}$$

with corresponding overall attachment rates

$$\pi_\pm(n_+, n_-) = k_\pm^{\text{on}}(N_\pm - n_\pm). \tag{6}$$

In which, the cargo load $F_C$ can be obtained by Eqs. (2) (3) (4). For *plus-motors winning cases,* i.e., $n_+ F_S^+ > n_- F_S^-$, one can verify

$$F_C(n_+, n_-) = \frac{V_F^+ + V_B^-}{V_F^+/n_+ F_S^+ + V_B^-/n_- F_S^-}. \tag{7}$$

So the detachment rate is [14]

$$k_\pm^{\text{off}}(n_+, n_-) = k_{0\pm}^{\text{off}} \exp\left[n_\mp/(an_+ + bn_-)F_d^\pm\right], \tag{8}$$

where the load distribution parameters are

$$a = \frac{V_B^-}{F_s^-(V_F^+ + V_B^-)}, \qquad b = \frac{V_F^+}{F_s^+(V_F^+ + V_B^-)}. \tag{9}$$

For discussing the mean-field limit for large $N_+$ and $N_-$ we set

$$y = n_+/N_+, \qquad z = n_-/N_-, \qquad \nu = N_+/N_-, \tag{10}$$

and then have

$$k_+^{\text{off}}(y, z) = k_{0+}^{\text{off}} \exp\left(\frac{z}{(a\nu y + bz)F_d^+}\right), \qquad k_-^{\text{off}}(y, z) = k_{0-}^{\text{off}} \exp\left(\frac{\nu y}{(a\nu y + bz)F_d^-}\right). \tag{11}$$

Similarly, for *minus-motors winning cases,* i.e., $n_+ F_S^+ < n_- F_S^-$, one can also get $k_\pm^{\text{off}}(y, z)$ by the above formulations, but with

$$a = \frac{V_F^-}{F_s^-(V_B^+ + V_F^-)}, \qquad b = \frac{V_B^+}{F_s^+(V_B^+ + V_F^-)}. \tag{12}$$

In the limit when $N_+$ and $N_-$ become large, the variables y and z become continuous and we obtain [14] the deterministic flow equations,

$$\begin{cases} \dfrac{dy}{dt} = k_+^{\text{on}} - y[k_+^{\text{on}} + k_+^{\text{off}}(y,z)] =: f(y,z), \\ \dfrac{dz}{dt} = k_-^{\text{on}} - z[k_-^{\text{on}} + k_-^{\text{off}}(y,z)] =: g(y,z), \end{cases} \quad (13)$$

where $k_\pm^{\text{off}}(y,z)$ can be obtained by Eq. (11) with

$$a = \frac{V_B^-}{F_s^-(V_F^+ + V_B^-)}, \qquad b = \frac{V_F^+}{F_s^+(V_F^+ + V_B^-)}, \quad \text{for } z \leq y\nu F_s^+/F_s^-; \quad (14)$$

$$a = \frac{V_F^-}{F_s^-(V_B^+ + V_F^-)}, \qquad b = \frac{V_B^+}{F_s^+(V_B^+ + V_F^-)}, \quad \text{for } z > y\nu F_s^+/F_s^-. \quad (15)$$

The corresponding velocity of cargo at state $(y,z)$ is

$$V_c(y,z) = \frac{y\nu F_s^+ - zF_s^-}{y\nu F_s^+/V_F^+ + zF_s^-/V_B^-}, \quad \text{for } z \leq y\nu F_s^+/F_s^-; \quad (16)$$

$$V_c(y,z) = \frac{y\nu F_s^+ - zF_s^-}{y\nu F_s^+/V_B^+ + zF_s^-/V_F^-}, \quad \text{for } z > y\nu F_s^+/F_s^-. \quad (17)$$

The cargo force is

$$F_c(y,z) = \frac{yz(V_B^- + V_F^+)N_+}{y\nu V_B^-/F_s^- + zV_F^+/F_s^+}, \quad \text{for } z \leq y\nu F_s^+/F_s^-; \quad (18)$$

$$F_c(y,z) = \frac{yz(V_B^+ + V_F^-)N_+}{y\nu V_F^-/F_s^- + zV_B^+/F_s^+}, \quad \text{for } z > y\nu F_s^+/F_s^-. \quad (19)$$

It can be easily verified that $f(y,z)$, $g(y,z)$, $k_\pm^{\text{off}}(y,z)$, $V_c(y,z)$ and $F_c(y,z)$ are all continuous functions. For $z = y\nu F_s^+/F_s^-$, i.e., $n_+ F_s^+ = n_- F_s^-$, $V_c(y,z) = 0$, $F_c(y,z) = n_+ F_s^+ = n_- F_s^-$, $k_\pm^{\text{off}}(y,z) = k_{0\pm}^{\text{off}} \exp\left(F_s^\pm/F_d^\pm\right)$.

### Stationary States and Their Stabilities

As discussed in [14, 15], one of the most interesting biological quantities of the tug-of-war model is its steady states $(y^*, z^*)$, which can be obtained by Eq. (13), i.e., $f(y^*, z^*) = 0$ and $g(y^*, z^*) = 0$. $(y^*, z^*)$ is stable if and only if the real parts of the two eigenvalues of the following Jacobi matrix are all negative

$$\begin{bmatrix} \frac{\partial f}{\partial y}(y^*, z^*) & \frac{\partial f}{\partial z}(y^*, z^*) \\ \frac{\partial g}{\partial y}(y^*, z^*) & \frac{\partial g}{\partial z}(y^*, z^*) \end{bmatrix}, \quad (20)$$

or equivalently,
$$\frac{\partial f}{\partial y}(y^*, z^*) + \frac{\partial g}{\partial z}(y^*, z^*) < 0,$$
$$\frac{\partial f}{\partial y}(y^*, z^*)\frac{\partial g}{\partial z}(y^*, z^*) - \frac{\partial f}{\partial z}(y^*, z^*)\frac{\partial g}{\partial y}(y^*, z^*) > 0. \tag{21}$$

From $f(y^*, z^*) = 0$ and $g(y^*, z^*) = 0$, one can see that $k_+^{\text{off}}(y^*, z^*) = (1-y^*)k_+^{\text{on}}/y^*$ and $k_-^{\text{off}}(y^*, z^*) = (1-z^*)k_-^{\text{on}}/z^*$, so it can be easily verified that

$$\frac{\partial f}{\partial y}(y^*, z^*) + \frac{\partial g}{\partial z}(y^*, z^*) = \frac{(1-y^*)z^* a\nu k_+^{\text{on}}}{(a\nu y^* + bz^*)^2 F_d^+} + \frac{(1-z^*)y^* b\nu k_-^{\text{on}}}{(a\nu y^* + bz^*)^2 F_d^-} - \frac{k_+^{\text{on}}}{y^*} - \frac{k_-^{\text{on}}}{z^*}, \tag{22}$$

and
$$\frac{\partial f}{\partial y}(y^*, z^*)\frac{\partial g}{\partial z}(y^*, z^*) - \frac{\partial f}{\partial z}(y^*, z^*)\frac{\partial g}{\partial y}(y^*, z^*)$$
$$= \left(\frac{1}{y^* z^*} - \frac{(1-z^*)b\nu}{(a\nu y^* + bz^*)^2 F_d^-} - \frac{(1-y^*)a\nu}{(a\nu y^* + bz^*)^2 F_d^+}\right) k_+^{\text{on}} k_-^{\text{on}}. \tag{23}$$

If $\frac{\partial f}{\partial y}(y^*, z^*)\frac{\partial g}{\partial z}(y^*, z^*) - \frac{\partial f}{\partial z}(y^*, z^*)\frac{\partial g}{\partial y}(y^*, z^*) \geq 0$, then from (23) one can see

$$\frac{(1-z^*)b\nu}{(a\nu y^* + bz^*)^2 F_d^-} \leq \frac{1}{y^* z^*}, \quad \frac{(1-y^*)a\nu}{(a\nu y^* + bz^*)^2 F_d^+} \leq \frac{1}{y^* z^*}, \tag{24}$$

and at least one of the inequalities holds strictly. So Eq. (22) implies $\frac{\partial f}{\partial y}(y^*, z^*) + \frac{\partial g}{\partial z}(y^*, z^*) < 0$. Thus the conditions (21) that stationary state $(y^*, z^*)$ is stable can be simplified as follows

$$\frac{\partial f}{\partial y}(y^*, z^*)\frac{\partial g}{\partial z}(y^*, z^*) - \frac{\partial f}{\partial z}(y^*, z^*)\frac{\partial g}{\partial y}(y^*, z^*) > 0, \tag{25}$$

or
$$\frac{1}{y^* z^*} - \frac{(1-z^*)b\nu}{(a\nu y^* + bz^*)^2 F_d^-} - \frac{(1-y^*)a\nu}{(a\nu y^* + bz^*)^2 F_d^+} > 0. \tag{26}$$

On the contrary, if $\frac{\partial f}{\partial y}(y^*, z^*)\frac{\partial g}{\partial z}(y^*, z^*) - \frac{\partial f}{\partial z}(y^*, z^*)\frac{\partial g}{\partial y}(y^*, z^*) < 0$, then stationary state $(y^*, z^*)$ is not stable. It should be pointed out that, if $\frac{\partial f}{\partial y}(y^*, z^*)\frac{\partial g}{\partial z}(y^*, z^*) - \frac{\partial f}{\partial z}(y^*, z^*)\frac{\partial g}{\partial y}(y^*, z^*) = 0$, then one eigenvalue of Jacobi matrix in (20) is negative, and another one is zero (note: in these cases $\frac{\partial f}{\partial y}(y^*, z^*) + \frac{\partial g}{\partial z}(y^*, z^*) < 0$). It means that bifurcation occurs at the corresponding parameters of the model, which might be transcritical bifurcation, saddle-node bifurcation, or pitchfork bifurcation [15]. Obviously, the stationary states near axes $y$ or $z$, i.e. $y^* \to 0$ or $z^* \to 0$, is stable.

## III. SIMPLIFICATION TO ONE-DIMENSIONAL EQUATION

It is easy to see that there is no steady state $(y^*, z^*)$ of Eq. (13) which lies on axis $y$ or $z$. Let $\theta = y^*/z^*$, i.e., $z^* = y^*/\theta$, then

$$\begin{cases} y^*[k_+^{\mathrm{on}} + k_+^{\mathrm{off}}(y^*, y^*/\theta)] = k_+^{\mathrm{on}}, \\ y^*[k_-^{\mathrm{on}} + k_-^{\mathrm{off}}(y^*, y^*/\theta)] = \theta k_-^{\mathrm{on}}. \end{cases} \qquad (27)$$

So the variable $\theta$ satisfies

$$h(\theta) := (\theta - 1)k_+^{\mathrm{on}}k_-^{\mathrm{on}} + \theta k_-^{\mathrm{on}} k_+^{\mathrm{off}}(\theta) - k_+^{\mathrm{on}} k_-^{\mathrm{off}}(\theta) = 0, \qquad (28)$$

where

$$k_+^{\mathrm{off}}(\theta) = k_{0+}^{\mathrm{off}} \exp\left(\frac{1}{(a\nu\theta + b)F_d^+}\right), \qquad k_-^{\mathrm{off}}(\theta) = k_{0-}^{\mathrm{off}} \exp\left(\frac{\nu\theta}{(a\nu\theta + b)F_d^-}\right), \qquad (29)$$

and

$$a = \frac{V_B^-}{F_s^-(V_F^+ + V_B^-)}, \qquad b = \frac{V_F^+}{F_s^+(V_F^+ + V_B^-)}, \qquad \text{for } \theta \geq \frac{F_s^-}{\nu F_s^+}, \qquad (30)$$

$$a = \frac{V_F^-}{F_s^-(V_B^+ + V_F^-)}, \qquad b = \frac{V_B^+}{F_s^+(V_B^+ + V_F^-)}, \qquad \text{for } 0 \leq \theta < \frac{F_s^-}{\nu F_s^+}. \qquad (31)$$

Therefore, $(y^*, z^*)$ is a steady state of Eq. (13) if and only if $\theta = y^*/z^*$ is one of the roots of Eq. (28). If $\theta^*$ is one of the roots of Eq. (28), then the corresponding steady state $(y^*, z^*)$ can be obtained as follows (see Eq. (27))

$$y^* = \frac{k_+^{\mathrm{on}}}{k_+^{\mathrm{on}} + k_+^{\mathrm{off}}(\theta^*)}, \qquad z^* = \frac{k_-^{\mathrm{on}}}{k_-^{\mathrm{on}} + k_-^{\mathrm{off}}(\theta^*)}. \qquad (32)$$

The corresponding velocity of cargo at steady state $(y^*, z^*)$ is

$$V_c(y^*, z^*) = V_c(\theta^*) = \frac{\theta^* \nu F_s^+ - F_s^-}{\theta^* \nu F_s^+/V_F^+ + F_s^-/V_B^-}, \qquad \text{for } \theta^* \geq \frac{F_s^-}{\nu F_s^+}, \qquad (33)$$

$$V_c(y^*, z^*) = V_c(\theta^*) = \frac{\theta^* \nu F_s^+ - F_s^-}{\theta^* \nu F_s^+/V_B^+ + F_s^-/V_F^-}, \qquad \text{for } \theta^* < \frac{F_s^-}{\nu F_s^+}. \qquad (34)$$

One can verify that $h(\theta)$ is a continuous function and

$$\begin{aligned} h(0) &= -k_+^{\mathrm{on}}(k_-^{\mathrm{on}} + k_{0-}^{\mathrm{off}}) < 0, \\ \lim_{\theta \to \infty} \frac{h(\theta)}{\theta} &= k_-^{\mathrm{on}}(k_+^{\mathrm{on}} + k_{0+}^{\mathrm{off}}) > 0. \end{aligned} \qquad (35)$$

More importantly, if $\theta^*$ satisfies $h(\theta^*) = 0$, and $y^*, z^*$ are obtained by formulation (32), then

$$\begin{aligned}\frac{dh(\theta^*)}{d\theta} &= k_-^{on}(k_+^{on} + k_+^{off}(\theta^*)) - \frac{b\nu k_-^{on}[(\theta^* - 1)k_+^{on} + \theta^* k_+^{off}(\theta^*)]}{(a\nu\theta^* + b)^2 F_d^-} - \frac{a\theta^*\nu k_-^{on} k_+^{off}(\theta^*)}{(a\nu\theta^* + b)^2 F_d^+} \\ &= \left(\frac{1}{y^* z^*} - \frac{(1-z^*)b\nu}{(a\nu y^* + bz^*)^2 F_d^-} - \frac{(1-y^*)a\nu}{(a\nu y^* + bz^*)^2 F_d^+}\right) z^* k_+^{on} k_-^{on} \\ &= \left(\frac{\partial f}{\partial y}(y^*, z^*)\frac{\partial g}{\partial z}(y^*, z^*) - \frac{\partial f}{\partial z}(y^*, z^*)\frac{\partial g}{\partial y}(y^*, z^*)\right) z^*.\end{aligned} \quad (36)$$

From (25) (36), one can see that $(y^*, z^*)$ is a stable (unstable) stationary state of system (13) is equivalent to that $\theta^* = y^*/z^*$ is one of the roots of $h(\theta) = 0$ with positive (negative) derivative. Bifurcation occurs at the parameters if there exists a root of $h(\theta) = 0$ with zero derivative.

For convenience of further analysis and numerical calculations, we define

$$w(\vartheta) = \begin{cases} h(\vartheta), & \text{for } 0 \leq \vartheta \leq 1, \\ (2-\vartheta)h\left(\frac{1}{2-\vartheta}\right), & \text{for } 1 < \vartheta < 2. \end{cases} \quad (37)$$

One can easily see that the zero points $\theta^*$ of $h(\theta) = 0$ ($0 \leq \theta < \infty$) and the zero points $\vartheta^*$ of $w(\vartheta) = 0$ ($0 \leq \vartheta < 2$) have the following relations

$$\theta^* = \begin{cases} \vartheta^*, & \text{if } 0 \leq \vartheta^* \leq 1, \\ \dfrac{1}{2-\vartheta^*}, & \text{if } 1 < \vartheta^* < 2. \end{cases} \quad (38)$$

Moreover, one can verify

$$w'(\vartheta) = \begin{cases} h'(\vartheta), & \text{for } 0 \leq \vartheta \leq 1, \\ \dfrac{\vartheta - 1}{2-\vartheta}h'\left(\dfrac{1}{2-\vartheta}\right), & \text{for } 1 < \vartheta < 2. \end{cases} \quad (39)$$

which implies, for $\theta^* \neq 1$,

$$\frac{dh(\theta^*)}{d\theta} > 0 \iff \frac{dw(\vartheta^*)}{d\vartheta} > 0, \quad (40)$$

It also can be easily verified that $w(\vartheta)$ is a continuous function in $[0, 2)$,

$$w(0) = h(0) = -k_+^{on}(k_-^{on} + k_{0-}^{off}) < 0 \qquad \lim_{\vartheta \to 2} w(\vartheta) = \lim_{\theta \to \infty} \frac{h(\theta)}{\theta} = k_-^{on}(k_+^{on} + k_{0+}^{off}) > 0. \quad (41)$$

Which implies that, there is at least one zero point $\vartheta^*$ which satisfies $dw(\vartheta^*)/d\vartheta > 0$. If there are more than one zeros $\vartheta_1^* < \cdots < \vartheta_k^*$, then $\vartheta_1^*, \vartheta_k^*$ will satisfy $dw(\vartheta^*)/d\vartheta > 0$. It is to say that the tug-of-war model has at least one stable motion mode. From Eq. (41), we also can see that, usually, the number of stationary states of the tug-of-war model is odd, and the ones which are the nearest to axes $y$ or $z$ are stable. The bifurcations will occur at the model parameters, at which the number of stationary states is even [15, 16], or in other words, there exists a root of $w(\vartheta) = 0$ with zero derivative. Since, in this cases, there exists one solution $\theta^*$ of Eq. (28) which satisfies $h(\theta^*) = h'(\theta^*) = 0$. For the sake of simplicity of the numerical calculations, the function $w(\vartheta)$ in Eq. (37) can be replaced by the following

$$\hat{w}(\vartheta) = \begin{cases} h(\vartheta), & \text{for } 0 \leq \vartheta \leq \dfrac{F_s^-}{\nu F_s^+}, \\ \dfrac{(2-\vartheta)}{2 - \frac{F_s^-}{\nu F_s^+}} h\left(\dfrac{2 - \frac{F_s^-}{\nu F_s^+}}{2 - \vartheta} \dfrac{F_s^-}{\nu F_s^+}\right), & \text{for } \dfrac{F_s^-}{\nu F_s^+} < \vartheta < 2. \end{cases} \quad (42)$$

The reason is that, the function $w(\vartheta)$ should be given different definitions in three different subintervals of $[0, 1)$, but $\hat{w}(\vartheta)$ only need to be defined in two subintervals of $[0, 1)$ respectively. It need to be pointed out here, although there are might two points in interval $[0, 2)$, $\vartheta = 1$ and $\vartheta = F_s^-/(\nu F_s^+)$ (if $F_s^-/(\nu F_s^+) \leq 1$) or $\vartheta = 2 - \nu F_s^+/F_s^-$ (if $F_s^-/(\nu F_s^+) > 1$), at which the derivative of $w(\vartheta)$ dose not exist, all the above results remain valid. If $\vartheta^*$ is one of such points, and $w'(\vartheta^*-)w'(\vartheta^*+) > 0$, we regard $w'(\vartheta^*)$ has the same sign as $w'(\vartheta^*-)$ and $w'(\vartheta^*+)$. Otherwise, we regard $w'(\vartheta^*) = 0$.

Finally, to illustrate the basic properties of function $w(\vartheta)$, and consequently the properties of tug-of-war model, the figures of $w(\vartheta)$ are plotted in Fig. 1. For the symmetric case, i.e., the plus-end and minus-end motors are different, the parameters used in the calculations are $k_\pm^{on} = k_{0\pm}^{off} = 1$ s$^{-1}$, $F_d^\pm = F_s^\pm = 1$ pN, $V_B^\pm = 10$ nm/s, and $V_F^\pm = 10, 30, 50$ nm/s respectively (Fig. 1 (left)); for the asymmetric case, $F_d^+$ and $F_s^+$ are changed to 1.2 pN, and $V_F^\pm = 20, 30, 40$ nm/s respectively (Fig. 1 (right)). As discussed above, the zero points of $w(\vartheta)$ with positive derivative correspond to stable stationary state of the model. Usually, the derivatives at the smallest and biggest roots of $w(\vartheta) = 0$ are positive. As the change of parameters, bifurcation might occur. At the corresponding parameters, the number of roots of $w(\vartheta) = 0$ is even. There exists $0 < \vartheta^* < 2$ which satisfies $w(\vartheta^*) = w'(\vartheta^*) = 0$.

## IV. CONCLUDING REMARKS

In this paper, the tug-of-war model for cargo transport in living cells is further studied theoretically. By changing variables, the two-dimensional model can be simplified into a one-dimensional equation. The stationary states of the tug-of-war model correspond to the zero points of the simplified equation. More importantly, we have mathematically proved that, the stability of the stationary states of the original model is determined by the sign of the derivative of a one dimensional function at the corresponding zero points. By the one dimensional model, we find that, usually the number of stationary states of the tug-of-war model is odd. The bifurcation occurs at the parameters, at which the number of stationary states is even, in which one of the stationary state corresponds to zero point of the simplified equation where the derivative of the one-dimensional function is zero. The stationary states of the tug-of-war model, which are nearest to the axes, $y$ or $z$, are usually stable, unless bifurcation occurs at the corresponding model parameters. Our simplification is also helpful to the numerical calculation of the tug-of-war model, using the one-dimensional equation, the accuracy and efficiency of the calculation will be greatly improved. Since the tug-of-war model is proved to be much successful to describe the cargo transport in cells [1, 3–5, 14, 15, 17, 18], the simplification in this paper will be very helpful to the future study about the corresponding biophysical problems.

### Acknowledgments

This work is funded by the National Natural Science Foundation of China (Grant No. 10701029).

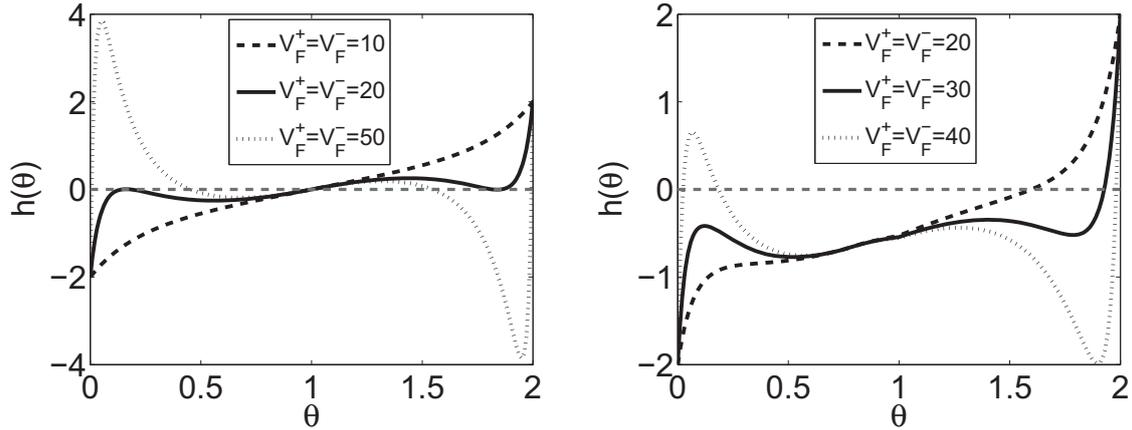

FIG. 1: The figures of function $w(\vartheta)$: (left) symmetric case, in which $k_{\pm}^{\text{on}} = k_{0\pm}^{\text{off}} = 1$ s$^{-1}$, $F_d^{\pm} = F_s^{\pm} = 1$ pN, $V_B^{\pm} = 10$ nm/s, and $V_F^{\pm} = 10, 30, 50$ nm/s respectively; (right) asymmetric case, in which $F_d^+$ and $F_s^+$ are changed to 1.2 pN, and $V_F^{\pm} = 20, 30, 40$ nm/s respectively. The zero points of $h(\theta)$ correspond to stationary states of the tug-of-war model, and the zero points with positive derivative correspond to stable stationary states.